\pdfoutput=1
\documentclass{osa-article}

\journal{oe}
\articletype{Research Article}
\usepackage{siunitx}
\DeclareSIUnit\ppm{\mathrm{ppm}}
\usepackage{hyperref}
\usepackage{lineno}
\usepackage{comment}
\begin{document}

\title{Laser written mirror profiles for open-access fiber Fabry-Pérot microcavities}

\author{Jannis Hessenauer,\authormark{1,*}, Ksenia Weber,\authormark{3},Julia Benedikter\authormark{4},Timo Gissibl,\authormark{3} Johannes Höfer,\authormark{1} Harald Gießen,\authormark{3} and David Hunger\authormark{1,2}}

\address{\authormark{1} Physikalisches Institut, Karlsruhe Institute of Technology (KIT), Wolfgang-Gaede-Str. 1, 76131 Karlsruhe, Germany\\
\authormark{2} Institute for Quantum Materials and Technologies (IQMT), Karlsruhe Institute of Technology (KIT), Hermann von Helmholtz Platz 1, Eggenstein-Leopoldshafen, Germany\\

\authormark{3}4th Physics Institute and Research Center SCoPE, University of Stuttgart, Pfaffenwaldring 57, 70569 Stuttgart, Germany\\

\authormark{4} Ludwig Maximilians University Munich, Schellingstr. 4, 80799 München, Germany\\
}
\email{\authormark{*}jannis.hessenauer@kit.edu} %% email address is required



\begin{abstract*}
We demonstrate laser-written concave hemispherical structures produced on the endfacets of optical fibers that serve as mirror substrates for tunable open-access microcavities. We achieve finesse values of up to 250, and a mostly constant performance across the entire stability range. This enables cavity operation also close to the stability limit, where a peak quality factor of  $1.5\times 10^4$ is reached. Together with a small mode waist of $\SI{2.3}{\micro \metre}$, the cavity achieves a Purcell factor of $C \sim 2.5$, which is useful for experiments that require good lateral optical access or otherwise large separation of the mirrors. Laser-written mirror profiles can be produced with a tremendous flexibility in shape and on various surfaces, opening new possibilities for microcavities.
\end{abstract*}

%%%%%%%%%%%%%%%%%%%%%%%%%%  body  %%%%%%%%%%%%%%%%%%%%%%%%%%
\section{Introduction}
Optical microcavities are a powerful tool for enhancing light-matter interactions with applications ranging from sensing \cite{yoshie2011optical,zhi2017single} and spectroscopy \cite{gagliardi2014cavity} to quantum optics \cite{vahala2003optical,Li2019}. Open-access Fabry–Pérot cavities built from microscopic concave mirrors \cite{li_tunable_2019,Pfeifer2022}  can combine a small mode cross section, high finesse, and open access. This offers the potential for strong enhancement of light-matter interactions, flexible sample introduction, and full tunability of spatial and spectral degrees of freedom. Realizing the cavity at the endfacet of an optical fiber enables direct fiber-to-cavity coupling without adding further complexity \cite{steinmetz_stable_2006,hunger_fiber_2010,Pfeifer2022}. A variety of methods has been developed to produce concave, near-spherical profiles as mirror substrates, including CO$_2$ laser machining \cite{greuter_small_2014,hunger_laser_2012,uphoff_frequency_2015,takahashi_novel_2014,muller_ultrahigh-finesse_2010,gallego_high-finesse_2016}, chemical etching \cite{trupke_microfabricated_2005,biedermann_ultrasmooth_2010,jin_micro-fabricated_2022}, focused ion beam milling \cite{dolan_femtoliter_2010,albrecht_narrow-band_2014,trichet_topographic_2015}, and thermal reflow \cite{cui_hemispherical_2006,roy_fabrication_2011}. A common aspect of these techniques is the top-down approach, i.e., the microstructuring of an existing material, which leads to certain constraints regarding which materials and geometries can be fabricated. In contrast, bottom-up fabrication by 3D direct laser writing (DLW) opens up a large flexibility in producing structures with very little constraints\cite{gissibl2016two,schell2014laser,smith2020three}, such that spherical profiles with almost any radius of curvature, profile depth, and diameter can be produced. For instance, hemispherical structures that cover an entire half space can be produced, which can be of interest to realize cavities that enhance one specific cavity mode and simultaneously suppress all other ones. Also, a hemispherical cavity can allow one to operate a cavity close to the stability limit, where the mode waist can approach the diffraction limit \cite{durak2014diffraction,nguyen2018operating}. This could allow for highest spatial resolution for scanning cavity microscopy and high Purcell factors despite large mirror separations. 
%Previous work with 3D printed photonic structures: 
A multitude of photonic structures have been realized via 3D laser printing, amongst them whispering gallery mode resonators \cite{liu2010direct,woska2020tunable,schell2013three}, light collecting devices such as solid immersion lenses \cite{colautti20203d}, photonic crystals \cite{von2010three,deubel2004direct}, optical waveguides \cite{colautti20203d,gao2020high}, as well as multi-lens objectives on the endfacet of optical fibers \cite{gissibl2016two}. 
First examples of monolithic Fabry-P\'erot cavities were demonstrated, although they were limited to very low quality  factors below 50\cite{ortiz2018fabrication,smith2020three}, due to the limited mirror reflectivity and surface quality caused by the finite distance between printed voxels.

Here, we optimize the writing process of spherical profiles on the endfacets of optical fibers and achieve hemispherical structures with a high rotational symmetry, accurate spherical shape, and sufficient surface quality to enable the operation of a Fabry-P\'erot cavity with a finesse of up to 250 and a quality factor of up to $1.5\times 10^4$. We characterize the cavity across the accessible stability range and observe a mode spectrum which follows closely the idealized prediction from Hermite-Gaussian modes. Together with the comparably constant Finesse up to  mirror separations close to the stability limit, this evidences the high shape quality and opens up the possibility for operation within the promising regime close to the stability limit.
As an example, we demonstrate scanning cavity microscopy of gold nanoparticles and observe a point spread function with a waist of $\SI{2.3}{\micro \metre}$, despite a rather large mirror radius of curvature of $\SI{80}{\micro \metre}$ and a cavity length of $\SI{70}{\micro \metre}$.

\section{Fabrication and Geometric Characterization}

\begin{figure}[htbp]
\centering\includegraphics[width=13cm]{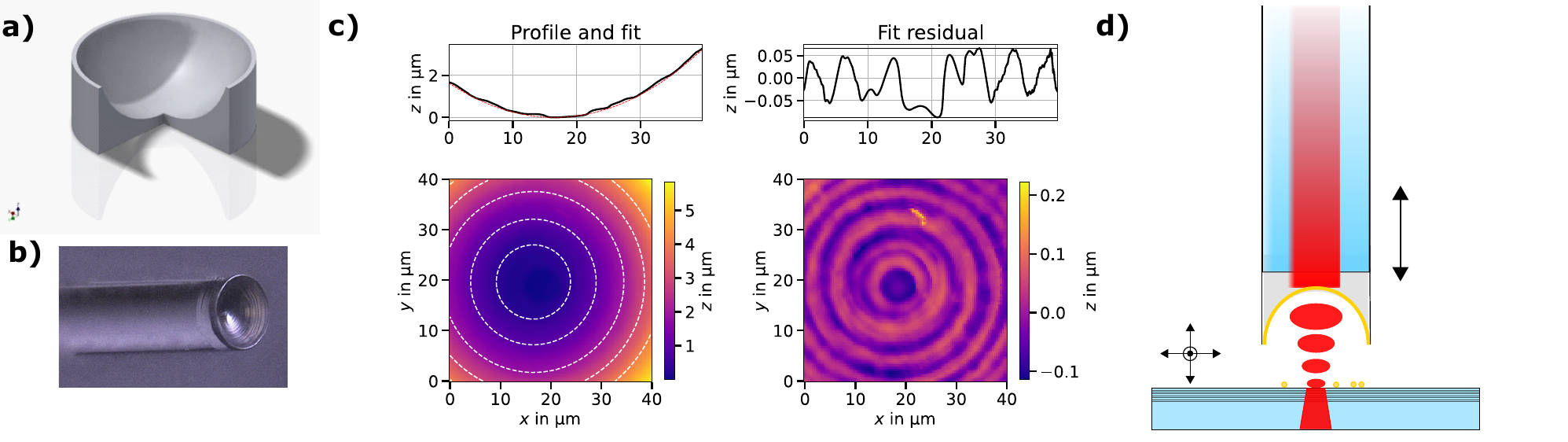}
\caption{ Characterization of the printed fiber profile.
a)	Schematic drawing of the printed concave profile.
b)	Micrograph of the printed profile on the endfacet of a single-mode fiber.
c)	Surface reconstruction of the printed profile of Fiber 1 measured with a white light interferometer.  The profile is well fit by a 2D-sphere yielding a radius of curvature of $r_\mathrm{c} = \SI{84}{\micro \metre}$. The residual reveals a wavy structure on a \SI{100}{\nano \metre} scale originating from the voxel lines in the printing process. 
d) Schematic illustration of the plano-concave scanning cavity geometry used in this work. 
}
\label{fig:1profile_characterization}
\end{figure}

The targeted fiber mirror geometry is a spherical profile with radius of curvature $r_c =\SI{78}{\micro \metre}$ on top of a small pedestal (Fig. \ref{fig:1profile_characterization} a)).   The structure was fabricated on the endfacet of a single mode fiber (HI 1060 Flex) via femtosecond two-photon direct laser lithography using a commercially available 3D direct laser printer (Photonic Professional GT, Nanoscribe GmbH). The fiber was mounted in the printer using a homebuilt fiber holder.
An optical micrograph of the structure is depicted in Fig. 1 b). After fabrication, the profile was coated with a $\SI{50}{\nano \metre}$ thick gold film via electron beam evaporation.
We characterize a typical profile by reconstructing the surface topography from white-light interferometric data taken in a homebuilt setup. Due to the limited depth of focus and range of resolvable surface angles of the microscope, we can not reconstruct the entire profile, but limit the imaged area to the central region ($40 \times 40 \,\si{\square \micro \metre}$). This is sufficient because the mode waist on the curved mirror does not exceed $\SI{15}{\micro \metre}$ even for our measurements close to the stability limit. The reconstructed profile is well fitted by a sphere, yielding a radius of curvature of $r_\mathrm{c}=\SI{84}{\micro \metre}$ and indicating a good overall match with the designed geometry. The reconstructed profile along with iso-contour lines from the fit and the fit residual are depicted in Fig. \ref{fig:1profile_characterization} c). The residual shows a wavy structure with a period of $4-5\,\si{\micro \metre}$ and deviations of roughly $\SI{100}{\nano \metre}$ peak to peak. This reflects the movement of the laser focus through the resist during the DLW fabrication. The laser spot was moved in a layer-by-layer fashion with the vertical steps between layers now showing up in the interferogram as concentric circles. This surface texture was consistently found for multiple produced fibers. Since the variations have a lateral length scale of a several micrometers, their effect on a cavity mode with a comparable size can not be described in the limit of Rayleigh scattering loss. Rather, the structures will lead to a distorted mode shape and mode mixing \cite{klaassen_transverse_2005,benedikter2015transverse}.

\section{Cavity Characterization via Finesse Measurements}
We have characterized two different fibers (henceforth labeled fiber 1 and fiber 2) in two slightly different configurations. The experimental setups are quite similar, so that only the one pertaining to fiber 1 is described here as an example. The main difference is the used laser wavelength, $\SI{940}{\nano \metre}$ for Fiber 1 and  $\SI{780}{\nano \metre}$ for Fiber 2, as well as suitable macroscopic planar Bragg mirrors to form the cavity. The cavity loss is dominated by scattering, absorption, and transmission at the gold-coated 3D-printed fiber mirror. The maximum achievable finesse is expected to be higher for fiber 1 due to the longer probing wavelength and therefore lower scattering and absorption and thus higher reflectance of the gold mirror. 
\begin{figure}
\centering\includegraphics[width=12cm]{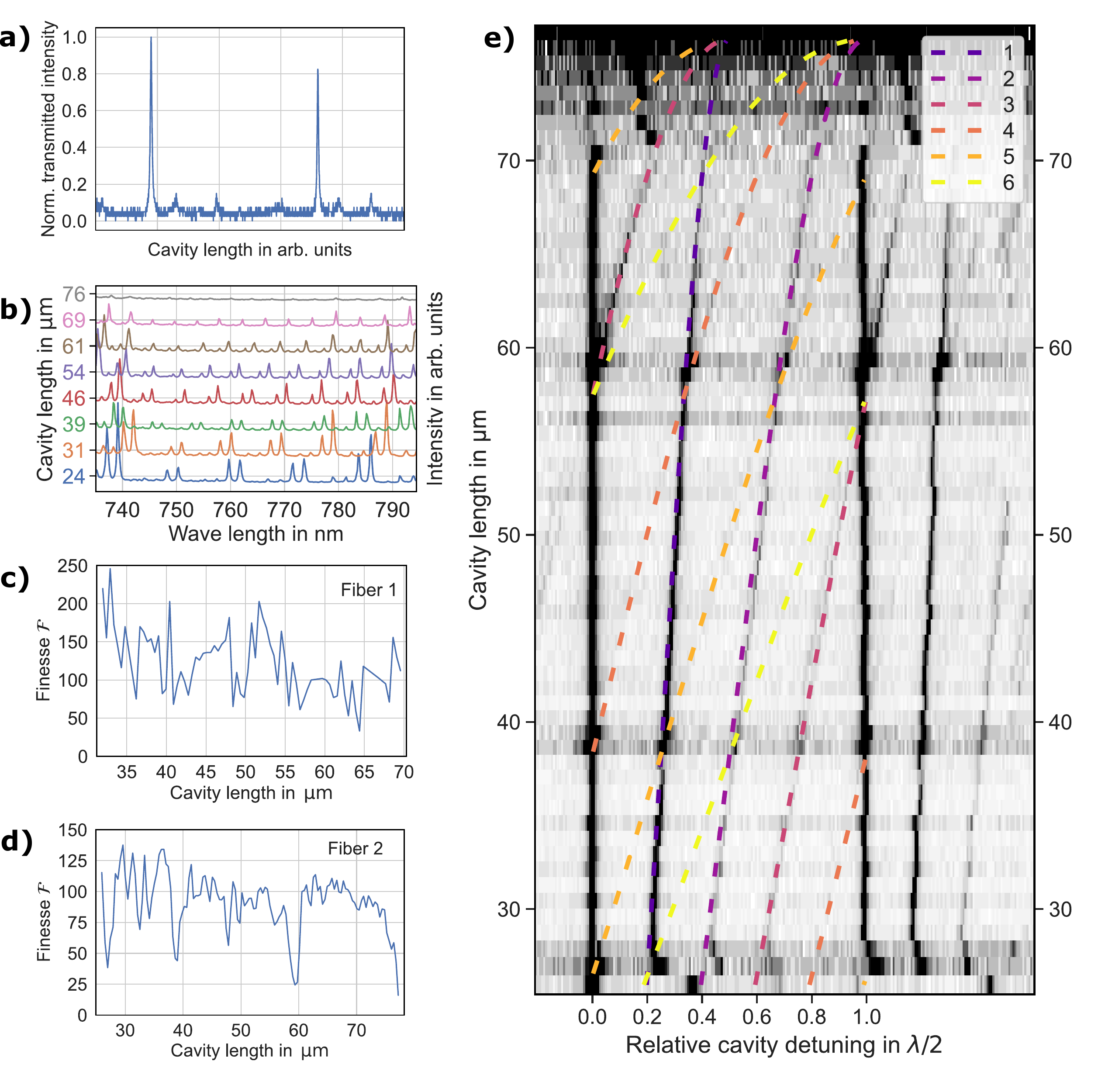}
\caption{%Cavity characterization of the printed mirror profiles
a) Typical cavity transmission spectrum used to calculate the finesse according to Eq. \ref{eq:Finesse}.
b) Cavity spectra for different cavity lengths measured by coupling a broadband light source into the cavity and detecting the transmission with a spectrometer. Spectra are offset for clarity.
c) Finesse measurement over the accessible stability range of Fiber 1. The finesse varies between 50 and 250.
d) Finesse measurement over the accessible stability range of Fiber 2. The finesse remains above 80 until approaching the stability limit. Pronounced dips in the finesse are observed, consistent with the points of resonant mode mixing.
d) Measured cavity transmission spectra over the entire accessible stability range of Fiber 2. Shifting higher order modes are clearly visible. The dashed lines are predictions of the position of the higher order transverse modes for a radius of curvature of $ r_\mathrm{c}=\SI{78}{\micro \metre}$, which was obtained from a measurement.}
\label{fig:2finesse_and_HO_modes}
\end{figure}
In order to characterize the mirror, we assemble a scanning micro-cavity from this fiber. The fiber is mounted inside a V-groove on a solid aluminium block. A planar half inch mirror with a distributed Bragg reflector coating centered at $\SI{985}{\nano \metre}$ is mounted in a custom holder that is fixed on a 3-axis nanopositioner (Attocube ECS3030) that enables lateral scanning of the mirror with respect to the fiber. The macroscopic mirror is approached to the fiber mirror to form the cavity. Light from an external cavity diode laser (Newport Velocity, $\lambda = \SI{940}{\nano \metre}$) is coupled into the fiber. The transmitted light is collected via a high NA objective and focused with a lens onto an avalanche photo diode and digitized with an oscilloscope (LeCroy Wavesurfer). In order to modulate the cavity length, a piezo-electric actuator driven by an amplified signal generator is used to move the fiber.
By scanning the cavity over one free spectral range, one can measure the line widths and the free spectral range of two succinct fundamental modes. The finesse $\mathcal{F}$ can then be calculated by the ratio of the two according to:

\begin{align}
    \mathcal{F} = \frac{\Delta_{\mathrm{FSR}}}{\Delta_{\mathrm{FWHM}}}
    \label{eq:Finesse}
\end{align}

A typical cavity transmission spectrum is shown in Fig. \ref{fig:2finesse_and_HO_modes} a). Notably, the higher order transverse mode families $\mathrm{TEM_{mn}}$ are degenerate for a fixed order $m + n$. This indicates a high rotational symmetry of the mirror profile and that the eigenmodes of the cavity are therefore Laguerre-Gaussian modes.
The cavity length $d$ can be measured by coupling a broadband light source into the cavity fiber and measuring the transmission spectrum with a spectrometer. Exemplary spectra for different cavity lengths are depicted in Fig. \ref{fig:2finesse_and_HO_modes} b). The cavity length can be calculated from the spectral position of two neighbouring fundamental modes, $\lambda_1$ and $\lambda_2$ as

\begin{align}
    d = \frac{\lambda_1 \lambda_2}{2(\lambda_2-\lambda_1)}
    \label{eq:CavityLength}
\end{align}

By moving the macroscopic mirror in z-direction, we can record the finesse as a function of the cavity length $d$. Fig. \ref{fig:2finesse_and_HO_modes} c) and d) show the finesse in the entire accessible stability range of the resonator for fiber 1 and 2, respectively. The mirror can not be approached closer than $d = \SI{32}{\micro \metre}$ ($d =\SI{23}{\micro \metre}$), limited by the heights of the printed profiles. As expected, we observe a stable finesse up to a cavity length of $d = \SI{70}{\micro \metre}$ ($d = \SI{75}{\micro \metre}$,), which approaches the stability limit $d = r_{\mathrm{c}}$. This is in contrast to typical $\mathrm{CO_2}$-laser machined mirror profiles, where in particular for small profiles, reduced finesse is observed for $d \gtrsim r_{\mathrm{c}}/2$ due to resonant mode mixing that originates from the non-spherical overall shape of such profiles \cite{benedikter2015transverse}.
The finesse remains above 100 for most of the stability range and reaches peak values of up to 250 (125) at small $d$. In particular for fiber 2, the finesse is mostly stable at high values and only shows a few distinct drops at certain cavity lengths. This is a typical behavior for resonant mode coupling, when the fundamental mode $\mathrm{TEM}_{00}$ resonantly couples to the first few higher order modes at particular mirror separations. This coupling is mediated by deviations from a perfect spherical mirror geometry, consistent with the residual printing lines on the printed mirror surface.

The transmittance and absorptance of the mirrors for the configuration used for Fiber 1 are calculated in a transfer matrix model to be $T_{\mathrm{Bragg}}  = \SI{400}{\ppm}$  ($T_{\mathrm{gold}}= \SI{0.93}{\%}$) and $A_{\mathrm{Bragg}}  = \SI{20}{\ppm}$ ($A_{\mathrm{Gold}}  = \SI{1.33}{\%}$). This limits the finesse to a value of $\mathcal{F}= 273$, without considering additional loss channels. The other dominant loss channel is scattering loss at the surfaces. The highest observed finesse of $\mathcal{F} = 250$ is consistent with scattering losses of $S = \SI{0.21}{\%}$. In a simple model, this corresponds to a RMS micro surface roughness of $\sigma_{\mathrm{sc}} = \frac{\lambda}{4 \pi}\sqrt{S}  \approx \SI{3}{\nano \metre}$. This is in accordance with the typical local micro roughness of evaporated gold films, and thereby confirms that the measured roughness of $\approx \SI{40}{\nano \metre}$ rms on a larger lateral length scale originating from the voxel lines does not contribute significantly here. 

Fig. \ref{fig:2finesse_and_HO_modes} e) shows the evolution of the cavity mode structure for Fiber 2 and how the distance between fundamental and higher order modes changes with increasing cavity length due to the additional Gouy phase for the higher order modes. The spectra agree well with the calculated resonance frequencies of Laguerre-Gaussian modes for a radius of curvature of $r_\mathrm{c} = \SI{78}{\micro \metre}$ as determined from white light interferometric measurements, which confirms the good spherical shape. 

\section{Enhancement of light-matter interactions}
The central figure of merit of a cavity for the enhancement of light-matter interactions is the cooperativity or similarly the Purcell factor, which can be expressed by the cavity quality factor $Q$ and mode volume $V$, or alternatively by the finesse $\mathcal{F}$ and beam waist $w_0$,

\begin{align*}
    C = \frac{3}{4\pi^2} \left( \frac{\lambda_0}{n} \right)^3 \frac{Q}{V} = \frac{6}{\pi^3} \left( \frac{\lambda_0}{n} \right)^3 \frac{\mathcal{F}}{w_0^2}.
\end{align*}

\begin{figure}
\centering\includegraphics[width=12cm]{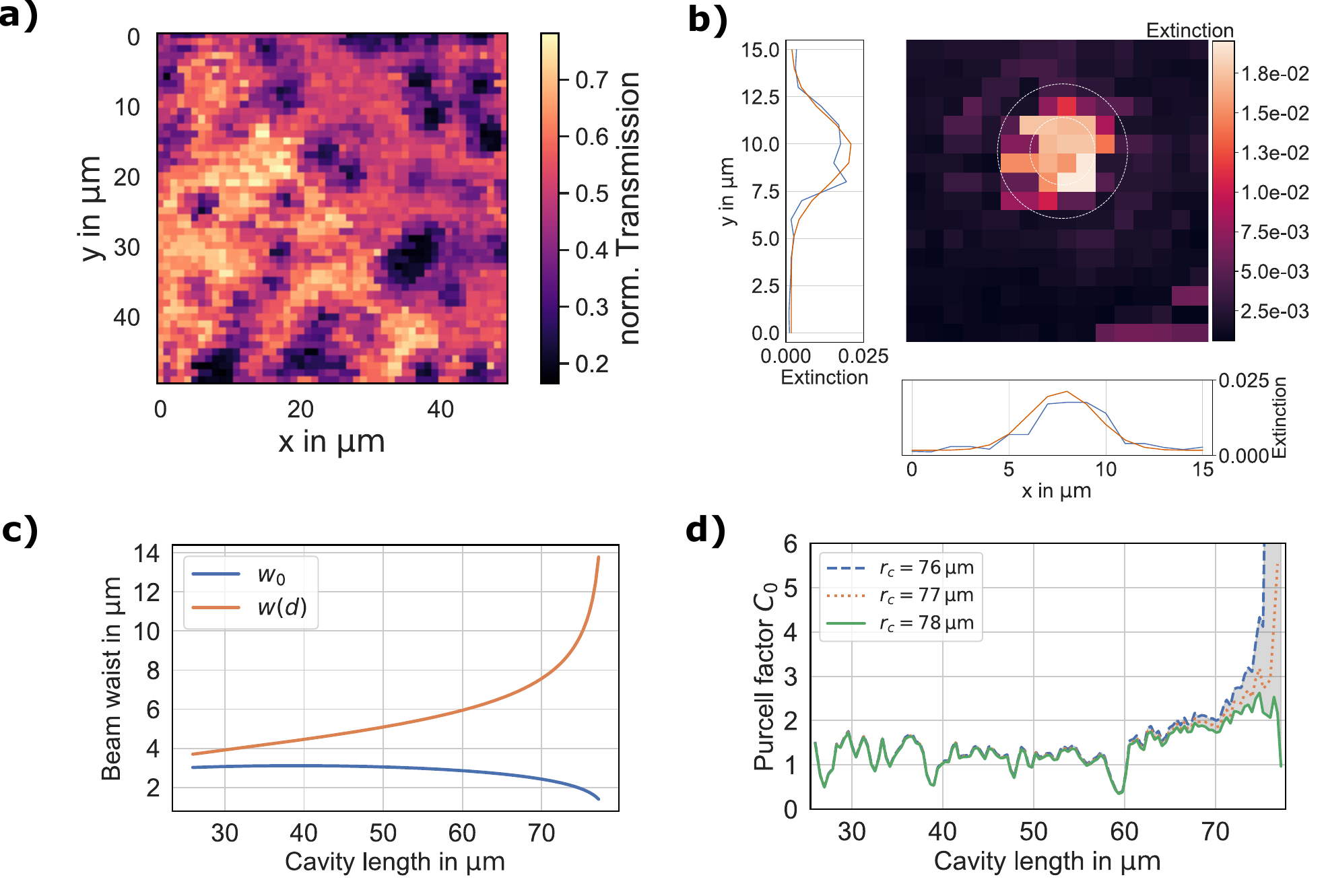}
\caption{
a) Scanning cavity transmission scan. Gold nanoparticles have been deposited on a dielectric mirror via dropcasting, forming large agglomerates with a circular shape as well as isolated particles which reveal the point spread function of the cavity. 
b) Extinction map of a single gold nanoparticle, showing a measured beam waist of $w_\mathrm{0} = \SI{2.35}{\micro \metre}$.
c) Calculated beam waist on both mirrors over the theoretical stability range of the cavity.
d) Expected Purcell factor calculated from the measured finesse and calculated beam waist.}
\label{fig:3_Beamwaist_Purcell}
\end{figure}

A minimal beam waist plays thus a central role, which is further beneficial to achieve a high spatial resolution in scanning cavity microscopy measurements. The beam waist $w_0$ of a plano-concave cavity is given by

\begin{align}
    w_0 = \sqrt{\frac{\lambda d}{\pi} \sqrt{\frac{r_\mathrm{c}}{d}-1}}.
    \label{eq:Beam waist}
\end{align}

The beam waist at the curved mirror is then obtained by propagating the Gaussian beam. Both quantities are plotted for Fiber 1 as a function of the cavity length $d$ in Fig. \ref{fig:3_Beamwaist_Purcell} c). One observes that the minimal waist $w_0$ remains comparably large over a large part of the stability range, but rapidly drops at very short and long cavity lengths close to the stability limit. These two regimes are therefore of great interest. Typically, the cavity length is reduced as much as possible to reduce the beam waist. The regime close to the stability limit so far remained elusive because of the imperfect, non-spherical mirror profiles created, e.g., by $\mathrm{CO_2}$ machining.

If the finesse $\mathcal{F}$ can experimentally be maintained independent of the cavity length, one can increase the light-matter interaction by operating the cavity close to the stability limit. The capability of the printed mirror structure is evidenced by the measurement in Fig. 2 d), where we observe for Fiber 2 a peak quality factor of $Q=1.5\times 10^4$ at a mirror separation $d=\SI{74}{\micro \metre}$. The expected Purcell enhancement is exemplary calculated from the measured finesse displayed in Fig. \ref{fig:2finesse_and_HO_modes} and the calculated mode waist in Fig.\ref{fig:3_Beamwaist_Purcell} c) and displayed in Fig. \ref{fig:3_Beamwaist_Purcell} d). The predicted Purcell factors at the stability limit are very sensitive to small uncertainties in the measurement of the radius of curvature and the cavity length. This is illustrated by plotting the curve for slightly different radii of curvature. Even for the lowest predicted values, the Purcell factor starts to increase at cavity lengths $d>\SI{60}{\micro \metre}$ up to a Purcell factor of $C \sim 2.5$ before the stability limit is reached.  This is a quite significant enhancement, considering the moderate finesse and large mirror separation.

To support the calculation of the beam waist with a measurement, we record the maximal transmission of a cavity mode at a given position and raster-scan the planar mirror to record extinction maps of the mirror surface, a technique called scanning cavity microscopy (Fig. \ref{fig:3_Beamwaist_Purcell} a)). This can be used to measure the beam waist of the cavity on the planar mirror. The straightforward approach is to measure the extinction profile of a pointlike scatter, which yields the shape of the recorded cavity mode as the point spread function. Therefore, we prepare a mirror with a sparse distribution of gold nanoparticles ($\varnothing  = \SI{60}{\nano \metre} << \lambda$) that act as point-like scatterers. By fitting the introduced loss map with a 2D-Gaussian profile, we can extract the beam waist $w_0$. An exemplary measurement is depicted in Fig. \ref{fig:3_Beamwaist_Purcell} b), yielding a beam waist of $w_\mathrm{0}  = \SI{2.37}{\micro \metre}$, where we average the radii of the semi-minor an semi-major axes of the beam waist. This value is in good agreement with the calculated beam waist.

\section{Conclusion}
Laser printing of hemispherical mirror profiles on the endfaces of optical fibers introduces a novel capability to realize open-access microcavities with increased flexibility. We have shown that despite residual surface structures from the printing process, a cavity finesse of up to 250 can be achieved that remains comparably constant at values $> 100$ until close to the stability limit $d=r_c$. At this point, a maximal Purcell factor of $\sim 2.5$ can be achieved, offering significant enhancement of light-matter interactions for a cavity with a large mirror separation. This relaxes the challenge to operate cavities at smallest mirror separation, allows one to introduce transversal laser beams, and enables the introduction of additional structures such as electrodes or microwave antennas. One can expect that significant improvements in surface quality are in reach with the latest developments in 3D laser printing. This can reduce scattering loss, and together with dielectric coatings, may enable high finesse cavities. This could open the way for high-finesse microcavities that can be integrated into complex, all-3D-printed structures for sensing, quantum optics or spectroscopy applications.

\begin{backmatter}
\bmsection{Acknowledgments}
This work has been financially supported by the Karlsruhe School of Optics and Photonics (KSOP), the Deutsche Forschungsgemeinschaft (DFG, German Research Foundation) under grant GRK 2643 Photonic Quantum Engineers, the Bundesministerium für Bildung und Forschung (BMBF) under the grants Q.Link.X (Contract No. 16KIS0879), QR.X (Contract No. 16KIS004), Printoptics, the European Research Counsil (ERC) (3DprintedOptics), the Carl-Zeiss-Stiftung (EndoPrint3D), the Gips-Schüle-Stiftung, and the Max-Planck-School of Photonics. 
\bmsection{Disclosures}
The authors declare no conflicts of interest.
\bmsection{Data Availability Statement}
Data underlying the results presented in this paper are not publicly available at this time but may be obtained from the authors upon reasonable request.
\end{backmatter}

\bibliography{LaserPrintedCavity}

\end{document}